\documentstyle[11pt,moriond,epsfig]{article}

\def\Journal#1#2#3#4{{#1} {\bf #2}, #3 (#4)}

\def\NCA{\em Nuovo Cimento}
\def\NIM{\em Nucl. Instrum. Methods}
\def\NIMA{{\em Nucl. Instrum. Methods} A}
\def\NPB{{\em Nucl. Phys.} B}
\def\PLB{{\em Phys. Lett.}  B}
\def\PRL{{\em Phys. Rev. Lett.} }
\def\PRD{{\em Phys. Rev.} D}

\def\EPC{{\em Eur. Phys.} C}

\def\be{\begin{equation}}
\def\ee{\end{equation}}
\def\bea{\begin{eqnarray}}
\def\eea{\end{eqnarray}}

\begin{document}
\vspace*{4cm}
\title{DIFFRACTIVE PRODUCTION OF THE HIGGS BOSON}

\author{R. PESCHANSKI}

\address{
CEA/DSM/SPhT, Unit\'e de recherche 
associ\'ee 
au CNRS, \\
CE-Saclay, F-91191 Gif-sur-Yvette Cedex, France\thanks{
	E-mail: {pesch@spht.saclay.cea.fr}}}

\maketitle\abstracts{Diffractive production of the Higgs boson at hadron 
colliders is discussed in the light of the observed rate of hard diffractive 
dijet events at the Tevatron. The Higgs predictions of models  successful for 
dijets are compared. LHC seems promising for  a diffractive light Higgs boson 
and its mass determination. Hard diffractive dijets, diphotons and dileptons at 
the Tevatron (Run II)  will be necessary to remove the remaining large 
uncertainties on cross-sections and signals.}

\section{Diffractive {\it vs.}
 Standard Higgs Production}

Standard Higgs boson production comes mainly from gluon-gluon fusion {\it via} 
the top quark loop, where the gluons are in the wave function of the incoming 
hadrons. The branching ratios are in $b \bar b,$ $\tau^+  \tau^-,$ and $\gamma 
\gamma,$ for the lower part of the Higgs boson mass spectrum to be  explored 
and $W^+ W^- ,$ $Z^0 Z^0,$ for masses above $160 \ GeV.$ The main problem to be 
faced by the Higgs boson hunt for low masses is that $b \bar b$ has an 
enormous 
QCD background,  $\tau^+  \tau^-,$ a difficult signature of decay modes and 
$\gamma \gamma,$ a small rate. While standard Higgs boson production is 
obviously the priority channel, it is thus useful to investigate  other  
channels, with smaller cross-sections but possibly cleaner signals.

Diffractive Higgs boson production provides such an  opportunity. Higgs boson 
production in the central rapidity region is mediated through color singlet 
exchanges leading to diffraction of the incoming hadrons at both vertices. 
Hence it is {\it a priori} possible to expect a single Higgs boson produced 
with large rapidity gaps separation with the diffracted hadrons (protons 
and/or 
antiprotons). Moreover, such diffractive processes can be selected 
by tagging 
initial  hadrons using  roman pot detectors, resulting in a pure signal: a Higgs 
boson and nothing else. 

Of course, this hope had to be supported by an  estimate of 
the cross-sections, which started with Ref.[1]. The difficulty with diffractive 
production is that it 
involves a mixture of perturbative and non-perturbative aspects of QCD which 
is far from a complete understanding. This has been reflected in the large
dispersion of predictions for the  diffractive Higgs boson production 
cross-sections during the 10 last years, preventing  reliable predictions for 
collider experiments.

\section{``Calibration'': Hard diffractive dijets  at 
Tevatron}

The recent  information allowing a progress in the determination of the 
cross-sections and the simulation of diffractive Higgs boson production came 
indirectly from the production of hard dijets observed \cite{cdf} at Tevatron. 
In the familiar configuration 
``Gap-Dijet-Gap'', it has been possible to register about 
100 dijet events with suitable cuts allowing to characterize them as created 
{\it via} double color singlet (the so-called DPE: ``Double Pomeron Exchange''). 

Let us introduce the appropriate kinematic variables, namely: ${\bf 
\xi_1(\xi_2)}$:   the $p (\bar p)$ 
fractional momentum loss (momentum fraction
carried by the Pomeron at each vertex); 
${\bf \Delta \eta_i \sim 
\log 1/ 
\xi_i}$: the rapidity gaps;
${\bf  M^2= s 
\xi_1\xi_2}$: The total (dijets plus soft hadronic radiation)
diffractive mass produced; 
${\bf  M^2_{JJ}}$: the dijet mass;
$ {\bf  {M^2_{JJ}} / {s \xi_1\xi_2}}$: The  dijet mass fraction. 

Interestingly enough (see Fig.1) the dijet mass fraction spectrum indicates 
the existence of a soft hadronic radiation accompanying the dijet system. 
Hence, the dijet diffractive production is neither essentially  ``exclusive'' 
(with just two jets) nor  ``soft'' (with a large gluon radiation coming from 
the incident particles). It can be called  ``inclusive''. Only a few models 
and a more restrictive range of predictions have been 
able to take into account this  peculiar experimental feature, and to describe 
the observed kinematical spectra and event rates.

\begin{figure} 
\begin{center} 
\includegraphics[width=5cm]{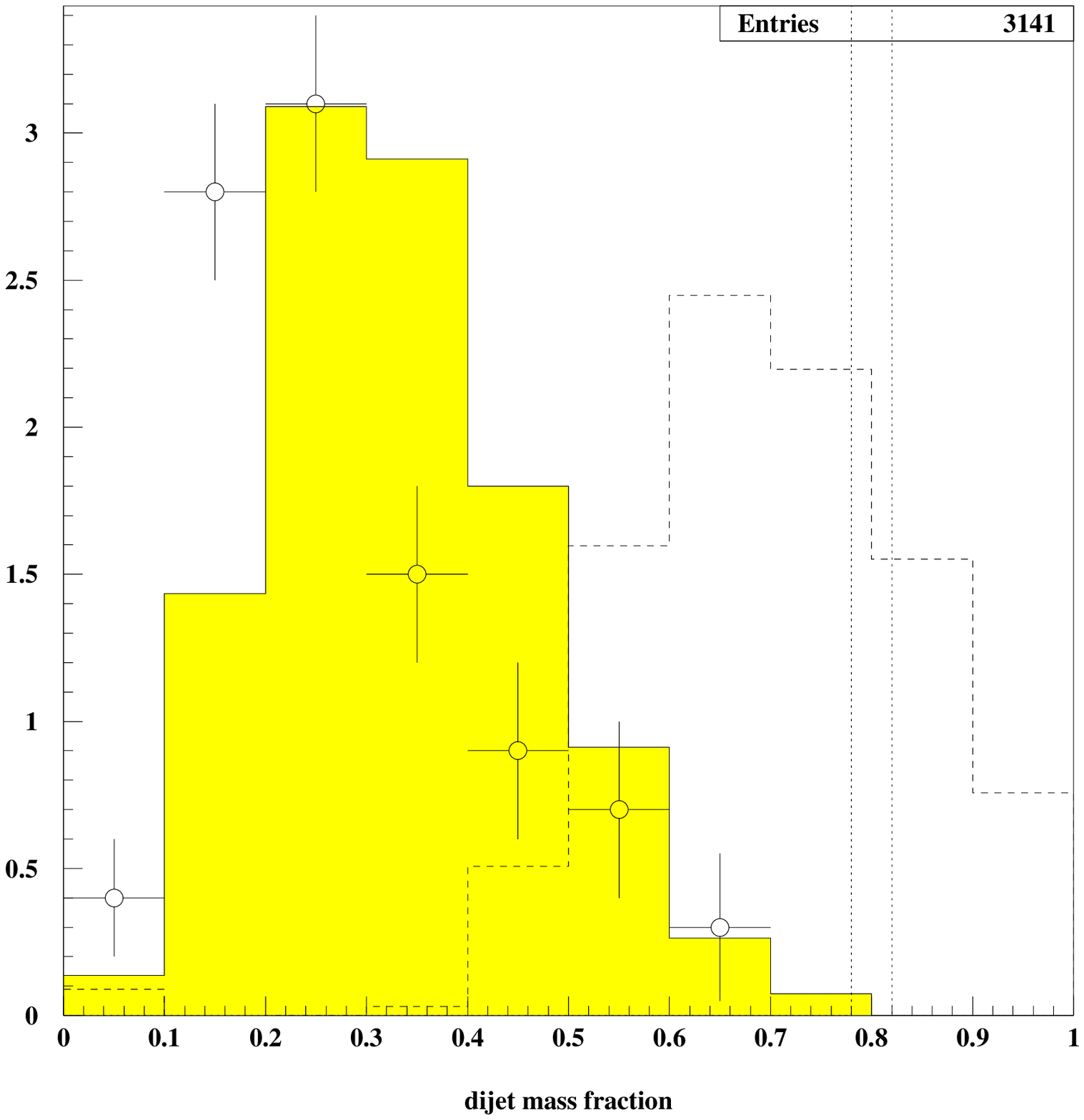} 
\includegraphics[width=5cm]{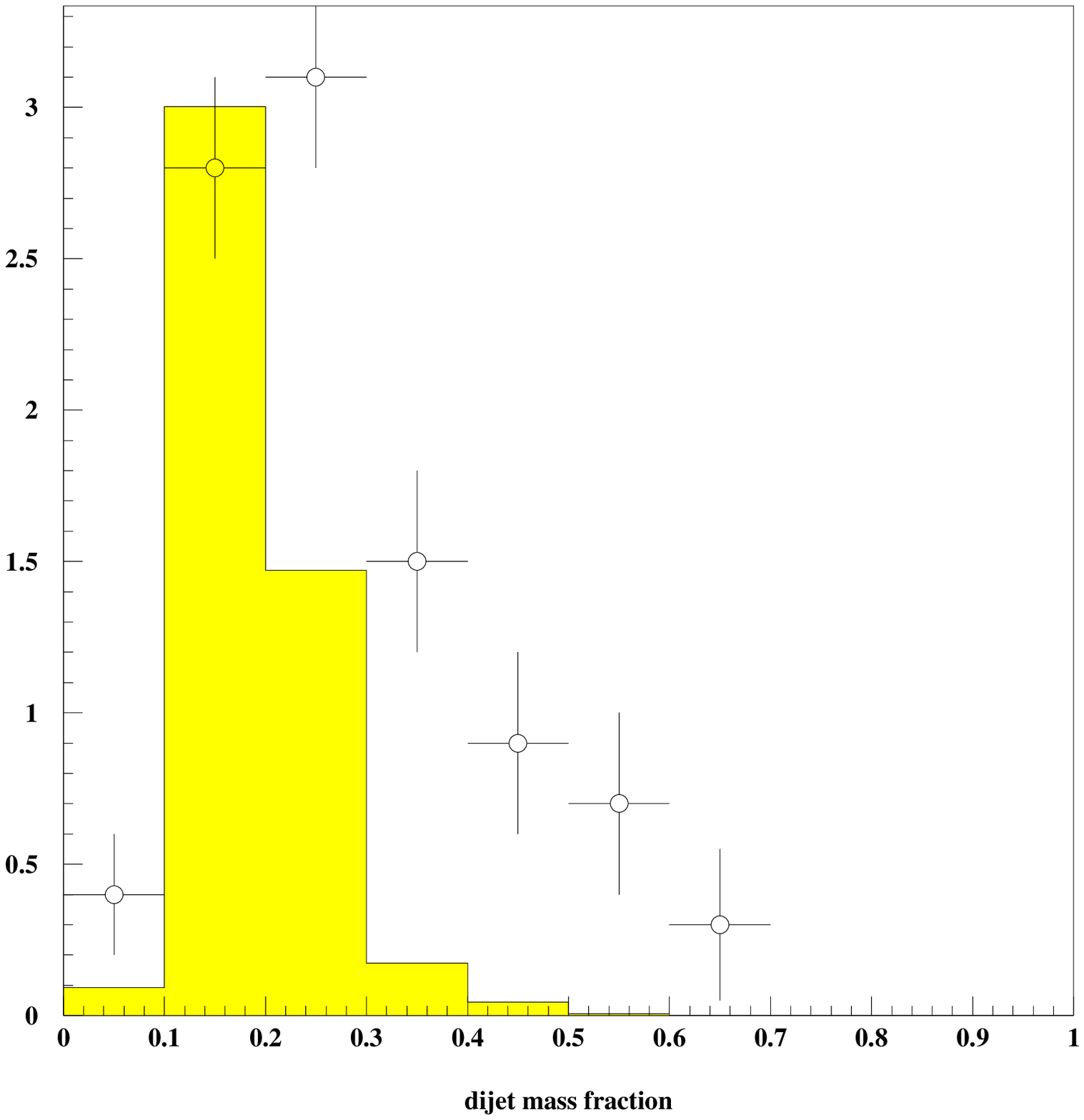} 
\caption{{\it Dijet mass fraction.} The data are from Ref.[2]. 
{\it Left:} (shaded) Gluons off the Pomeron (from model [3]); (white) 
``Exclusive'' mode,  before (column) or after (histogram) detector 
simulation.
{\it Right:} (shaded) Gluons off the proton.}
\end{center}
\label{fig:1}
\end{figure}

Among the list of models having been proposed for hard diffractive production 
in the central region, only four (types of) models seem to survive the 
compatibility  constraints with dijet production. Following the schemes of 
Fig.2, one has: 

{\bf (1)}  the non factorizable Pomeron model \cite{BPR};

{\bf (2)}  the factorizable Pomeron model \cite{CFH};

{\bf (3)}  The soft color interaction model \cite{SCI}.

{\bf (4)}  The exclusive model \cite{KMR}; 

The models {\bf (1,2)} describe hard dijet diffraction {\it via} the exchange of 
color singlets (Double Pomeron exchange). The ``inclusive'' dijet mass ratio are 
well described (see {\it e.g.}  model {\bf (1)}, Fig.1)  by the hadronic 
radiation associated with the Pomeron structure functions determined in hard 
diffraction processes at HERA. In model {\bf (2)}, the known HERA/Tevatron 
factorization breaking is due to the rapidity gap survival, while in  model {\bf 
(1)}, it is due to the soft gluon propagator as in [1], leading to  different 
event rate predictions. In the model {\bf (3)}, it is also described in the 
same way as diffraction at HERA, but using the soft color interaction formalism. 
Let us note the remaining  ``exclusive'' model 
{\bf (4)}, which event rate is sufficiently damped by the gap survival 
probability to stay below the present experimental limit for the exclusive 
cross-section. We shall focus on the discussion of Higgs boson 
production on the predictions of these four models which are able to describe or 
at least be compatible with the present data.

\begin{figure} 
\begin{center}
\includegraphics[width=8cm]{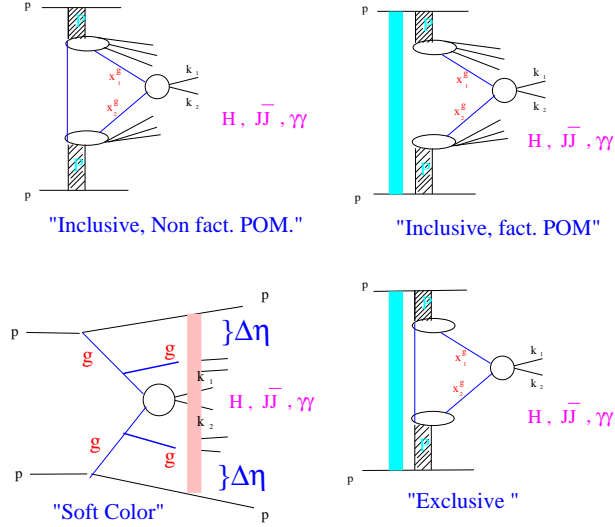}
\caption{{\it Models of hard diffractive production.} The shaded bands describe 
soft correction factors. For the
initial state: {\it  Gap Survival Probability}. For the final state: {\it  Soft 
Color Interaction.}}
 \end{center}
\end{figure}

\section{Predictions  for the  Tevatron and 
LHC}
Considering the four selected models, it is possible to check (or determine) 
their overall normalization to  the dijet event rate. Hence, by assuming a safe  
extrapolation to the Higgs boson cross-sections, one gets a range of predictions 
for the Tevatron (run II) and the LHC which is summarized on Table 1.

\begin{table}[b]
\caption{Model predictions for diffractive Higgs boson cross-sections (in 
femtobarns) 
.\label{tab:model}}
\vspace{0.4cm}

\begin{center} \begin{tabular}{|c||c|c|c|c|}
\hline 
&{\bf (1)}&{\bf (2)}&{\bf (3)}&{\bf(4)} \\
\hline\hline H$\sim$115 GeV, TeV. & 
1.7 & 0.029 & 10$^{-4}$ & .03 \\
${\cal L} \sim 1 fb^{-1}$ &  &0.09 & & \\
\hline
H $\sim$115 GeV, LHC & 169 & 379$\times$ & 0.19 & 1.4  
\\
$\times = GSP\sim(.1\to.03)$ & &486$\times$ & & \\
\hline
H $\sim$160 GeV, LHC & 123 & 145$\times$ & - & .55 \\
\hline 
\end{tabular} \end{center}
\end{table}

The first line of Table 1 is for the Tevatron with a typical low Higgs 
mass. We see that all models predict too few events to be observable. On the 
contrary, three of the four models predict a significant cross-section (w.r.t. 
the expected luminosity) at the LHC, even in the presence of a small gap 
suppression factor (GSP). The soft color interaction model prediction {\bf (3)}, 
however, is still low, which is a incentive for further tests of the models. For 
completeness, the model predictions for a larger Higgs mass are given in the 
last line of the table.

Table  1 shows that the ``inclusive'' cross-sections are   predicted in 
Pomeron models to be 10 to 100 times larger at LHC than the ``exclusive'' ones. 
It could even be possible that the exclusive mode could be completely hidden by 
the inclusive one. However, the  ``inclusive''  mode is hindered by the 
accompanying hadrons, the so-called Pomeron remnants. Indeed, this radiation may 
obscure the Higgs boson signal and {\it e.g.} the mass determination. In a 
recent publication \cite{BDPR}, it was proposed to trigger on the Pomeron 
remnants, in order to improve the signal. In some sense, the idea is to find a 
compromise between the much higher cross-sections of the inclusive mode and the 
clean signal of the exclusive one. the improvement of the Higgs boson mass 
determination with the trigger on Pomeron remnants is shown in Fig.3. 

The present uncertainties on cross-sections and signals remain quite large. They 
are mainly due to the diffractive cuts, the jet energy scale and the questions 
about the  soft correction factors. Hopefully,  stringent tests and 
discrimination of models will be provided  by the high statistics of hard DPE 
dijets, diphotons and dileptons expected from  the Run II. The lepton over 
photon rate will be able to distinguish between Pomeron and non-Pomeron models 
and the dijet mass differential spectum, between factorizable and 
non-factorizable models \cite{BPR}.

\begin{figure} 
\begin{center} 
\includegraphics[width=8cm]{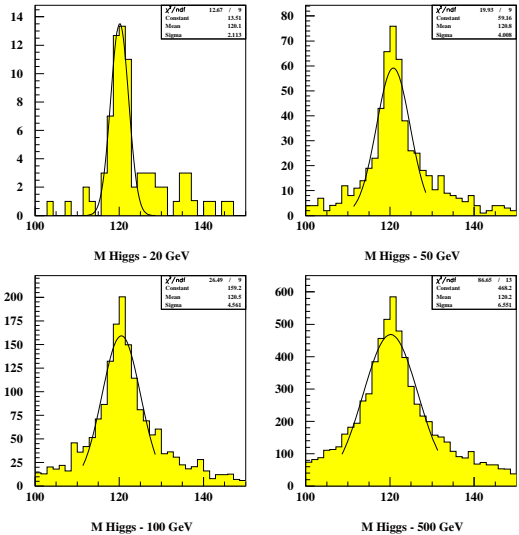}
 \caption{{\it Higgs mass reconstruction, from [7].} With a Higgs mass of $120\ 
GeV,$ the 
simulated  Higgs mass reconstruction is displayed with a cut of $20,\ 50,\ 100,\ 
500\ GeV$ on total Pomeron remnants' energy.} 
\end{center}
\end{figure}

\section*{Acknowledgments}
Thanks to M.Boonekamp and C.Royon, with whom the (non 
standard) Higgs hunting is a challenge and a pleasure and to  A.Bialas, A.De 
Roeck for collaboration.

\end{document}